\documentclass[aps,pre,preprint,superscriptaddress]{revtex4-1}
\usepackage{graphicx,calc,empheq}
\usepackage[T1]{fontenc} %en-able hy-phen-ation to op-er-ate on texts con-tain-ing any char-ac-ter in the font
\usepackage{amsmath,amsthm} % dovrebbero essere caricati di defould da aps, ma per sicurezza...
\usepackage{amssymb }  % Carica amsfonts piu altro
\usepackage{eucal}  % Euler script symbols in math mode. Comando: \mathcal{A} (anziche \CMcal{A})
\usepackage{hyperref}
\usepackage{enumitem}  % enu-mer-ate, item-ize and de-scrip-tion
\usepackage{bm}
\usepackage{dcolumn}% Align table columns on decimal point
\usepackage{color}

\newcommand{\N}{\mathbb{N}} 
 
\if@mathematic
   
\else
   
\fi

\if@vprodwedge
  
\else
  
\fi

\begin{document}

%\setpagewiselinenumbers
%\modulolinenumbers[5]
%\runninglinenumbers
%\preprint{}

\title{Optimal FPE for non-linear 1d-SDE.\\
I: Additive Gaussian colored noise}
\author{Marco Bianucci}
\affiliation{Istituto di Scienze Marine, Consiglio Nazionale delle Ricerche (ISMAR - CNR),\\
Forte Santa Teresa, Pozzuolo di Lerici, 19032 Lerici (SP), Italy}

\author{Riccardo Mannella}
\affiliation{Dipartimento di Fisica, Universit\`a di Pisa, 56100 Pisa, Italy}

%\thanks{}
\date{\today}


\date{\today}
\begin{abstract}
Many complex phenomena occurring in physics,
chemistry, biology, finance, etc.~\cite{Schadschneider2011} can be reduced, by some projection process, 
to a 1-d stochastic Differential Equation (SDE) for the variable of interest. 
Typically, this SDE is both non-linear and non-markovian, so a Fokker Planck equation (FPE), for the probability density function (PDF), is generally not obtainable. 
However, a FPE is desirable because it is the main  
tool to obtain relevant analytical statistical
information such as  stationary PDF and  First Passage Time. 

This problem has been addressed by many authors in the past (see among others~\cite{lwlPRA37, tgPRA38, fPRA34}), but due to
an incorrect use of the interaction picture (the standard tool to obtain a reduced FPE) previous theoretical results were
incorrect, as confirmed by direct numerical simulation of the SDE.

We will show, in general, how to address the problem and we will derived the correct best FPE from a perturbation approach. The method followed and the results
obtained have a general validity beyond the simple case of exponentially
correlated Gaussian driving used here as an example; they can be applied even 
to non Gaussian drivings with a generic time correlation.

\end{abstract}

%\keywords{}

\maketitle

\section{Introduction}
%
%\nolinenumbers
%\linenumbers
In the present work we are interested in non-linear 1-d SDEs  of the form:
\begin{align}
\label{SDEGen}
\dot {X}= - C(X) +  \epsilon \xi (t).
\end{align}
where $X$ is the variable of interest, $-C(X)$ is the unperturbed drift field,
$\xi(t)$ is the stochastic Gaussian perturbation with zero mean and autocorrelation function 
$\varphi(t)=\langle \xi (0) \xi(t) \rangle/\langle\xi^2\rangle$, the parameter 
$\epsilon$ controls the intensity of the perturbation, and $\langle ... \rangle$
implies average over the $\xi$ realizations. 
The SDE in Eq.~(\ref{SDEGen}) is ubiquitous in many research fields~\cite{Schadschneider2011}.

The choice to consider the simple additive and Gaussian SDE is made because here we want to focus attention on a pitfalls that, in our opinion, must be considered and solved, when applying perturbation methods to dissipative systems. The extension to the case of multiplicative coloured noise, possible non-Gaussian, will be dealt with in a later work.

It is a standard result in statistical physics that when the stochastic forcing $\xi$ is a ``white noise'', 
$\langle \xi(t) \xi(t') \rangle=2\,\delta(t-t') $, 
% $\langle \eta(t) \eta(t') \rangle=2 D_{\eta} \,\delta (t-t')$ 
Eq.~(\ref{SDEGen}) leads to a flow for the Probability Density Function (PDF) $P(X;t)$ of the variable $X$  equivalent to the probability flow given by the following  Fokker Planck Equation (FPE) (where $\partial_X:=\partial/\partial X$, $D_0= \epsilon^2 .$):
\begin{align}
\label{FPEexact}
\partial_t P(X;t)&=\partial_X C(X)\,P(X;t) + D_0\,\partial_X^2 P(X;t).
\end{align}
%
%+D_{\eta}$.
From Eq.~(\ref{FPEexact}), the stationary PDF is given by
\begin{equation}
\label{Pwhite}
P_{W,eq}(X)=\frac{1}{Z}  e^{-\int^X \frac{C(y)}{D_0} \, \text{d}y}
\end{equation}
where $Z$ is a normalization constant.

However, white noise is often an oversimplification of the real driving acting on a system of interest. Correlated noise 
(termed ``colored'' in the literature) is more common in continuous systems, and its importance has
 been recognized in a number of very different  situations, like for instance
 statistical properties of dye lasers~\cite{zPRA34,zPRA47,wlbCTP11,mxcwAPS8}
chemical reaction rate~\cite{oPA257,fgpJCP83,bgJCP96,bgpJCP92}, optical bistability~\cite{lwcCRP17,hlNoiseInducedTransitions}, large 
scale Ocean/Atmosphere dynamics~\cite{jltzGRL07,bGRL43} and many others. 

We will assume that the stochastic process
$\xi(t)$ is characterized by a ``finite'' correlation time $\tau$~\footnote{ The general prescription is that  there is a 
time $\tau$ such that, for any
        time $t$, the instances of $\xi$ at times $t'>t+\tau$  are ``almost statistically uncorrelated'' with the instances of $\xi$ at  times $t'<t$. 
        For ``almost statistically uncorrelated'' we mean that the joint probability density functions factorizes up to terms $O(\tau)$: $p_n(\xi_1,t_1';\xi_2,t_2';...;\xi_k,t_k';\xi_{k+1},t_{1};...;\xi_n,t_h)=
        p_k(\xi_1,t_1';\xi_2,t_2';...;\xi_k,t_k')\,p_h(\xi_{k+1},t_{1};...;\xi_n,t_h)+O(\tau)$ with $k,h,n\in \N$,  $k+h=n$ and $t_i'> t_j+\tau$.
         For example, $p_2(\xi_1,t';\xi_2,t)=p_1(\xi_1,t')\,p_1(\xi_2,t)+O(\tau)$.}
and unitary intensity $\langle\xi^2\rangle \tau=1$.  
It is well known that if the unperturbed drift field is 
linear, regardless of the number of dimensions, the  Gaussian property  of a generic colored noise $\xi(t)$ 
is ``linearly'' transferred to the system of interest, 
so the FPE structure does not break (see, e.g.,~\cite{aJCP76,bgpJCP92}). 
On the contrary, in the case of non linear SDE and/or non Gaussian noise, for finite values of $\tau$ the FPE structure 
breaks down. 
This is the case of interest here, and the aim of this paper is to recover in 
some appropriate limits
a FPE structure, obtaining an effective FPE with 
state dependent diffusion coefficient:
\begin{align}
\label{FPEGen}
\partial_t P(X;t)&=\partial_X C(X)\,P(X;t) + \partial_{X^2}^2 D(X) P(X;t)
 \end{align}
that, with a good approximation, could describe the 
evolution and the stationary properties of $P(X;t)$.
Given $D(X)$, the stationary PDF of the FPE
% of Eqs.~(\ref{FP_F3})-(\ref{FP_F4})
of Eq.~(\ref{FPEGen}) is then easily obtained
%\begin{subequations}
\begin{equation}
\label{Ps}
%\begin{equation}
%\label{PsBFPE}
P_s(X)=\frac{1}{Z}\frac{e^{-\int^X \frac{C(Y)}{D(Y)} \text{d}Y}}{D(X)} 
\end{equation}

Several techniques have been developed to deal with the correlation time of the
noise in nonlinear SDE, with the aim of eventually obtaining this effective FPE.
They can be grouped in three main categories that correspond to three general techniques: 
the cumulant expansion technique~\cite{kuboGenCumJPSJ17,kuboGenCumJMP4,bJMP59},  
the functional-calculus approach~\cite{fPRA33,fPRA34} (see also~\cite{lwlPRA37}) and
the projection-perturbation methods~(e.g.,~\cite{grigo_memory,grigolini1989,bgJCP96,bJSTAT2015}).
Each of these methods leads to a formally exact evolution equation for the PDF of the 
 driven process, and
the different descriptions are therefore
equivalent. The exact formal results do not lend themselves
to calculations nor give a FPE structure, therefore they require that approximations be made. 
The approximations made within these various formalisms
involve truncations and/or partial resummations
of infinite power series with respect to $\epsilon$ and $\tau$, which
are typically the small parameters in the problem. 
Not surprisingly, it has been argued~\cite{lwlPRA37} that the effective FPE obtained from
the different techniques are identical, if the same approximations are made 
(time scale separation, weak perturbation, Gaussian noise etc.). 
The results of the approximations can be grouped in two categories:  the ``Best
Fokker Plank Equation'' (BFPE)
 obtained by Lopez, West and Lindenberg~\cite{lwlPRA37} 
from a standard perturbation method, where $\epsilon$ is the small parameter and 
$\tau$ is finite but not limited,
and the ``Local Linearization Assumption'' (LLA) FPE
of  Grigolini~\cite{tgPRA38} that coincides with the result of the functional-calculus approach of Fox~\cite{fPRA33,fPRA34}.

However, strangely enough, the BFPE often fails when compared
with numerical simulations, even for relatively weak perturbations, while the LLA FPE usually works better. In Section~\ref{sec:LLA}
we will  comment briefly on this, leaving a more in-depth discussion to a later work.
 
In section~\ref{sec:FPE} we will shortly review the perturbation approach that leads to the BFPE,  stressing that care must be taken when using  
the interaction picture in strongly dissipative systems: the pitfalls we will point out are the sources of the defects of the original formulation of the BFPE. 
Section~ \ref{sec:curedBFPE} is the main section of the present work: we will
show how to cure the shortcomings of the BFPE pointed out in section~\ref{sec:FPE}.
Section~\ref{sec:LLA} is devoted to a comparison with the LLA results. 
In section~\ref{sec:conclusion} we present the conclusions.%

\section{The standard BFPE}\label{sec:FPE}

From Eq.~(\ref{SDEGen}) it follows that, for any realization of the process $\xi(u)$, with $0\le u\le t$,  
the time-evolution of the PDF of the whole system, which we indicate with $P_{\xi}(X;t)$, satisfies the following PDE:
\begin{equation}
\label{p(t)}
\partial_t P_{\xi}(X;t)={\cal L}_a\,P_{\xi}(X;t)+ \epsilon\, \xi(t) {\cal L}_I\, P_{\xi}(X;t)
\end{equation}
in which the  unperturbed Liouville operator ${\cal L}_a$ is
\begin{equation}
\label{La}
{\cal L}_a:=\partial_X C(X)
\end{equation}
and the Liouville perturbation operator is
\begin{equation}
\label{LI}
{\cal L}_I:=\partial_X.
\end{equation}
A standard step of the perturbation method is to introduce the interaction representation, by which Eq.~(\ref{p(t)}) becomes
\begin{equation}  
\partial_t \tilde{P}_\xi (X;t) 
= \epsilon \xi(t)\,\tilde{\cal L}_I(t) \tilde{P}_\xi (X;t),  
\label{app:liouvilleEq.1}  
\end{equation}  
where 
\begin{equation} 
\tilde{P}_\xi (X;t):=  
e^{-{\cal L}_a t}P_\xi ( X;t),
\end{equation}
and
\begin{equation}
\label{LIt} 
\tilde{\cal L}_I(t) := e^{-{\cal L}_a t}{\cal L}_I e^{{\cal L}_a t}=e^{-{\cal L}_a^\times t} [{\cal L}_I],
\end{equation}
where, for any couple of operators ${\cal A}$ and ${\cal B}$, we have defined 
${\cal A}^\times[{\cal B}]:=[{\cal A},{\cal B}]={\cal A} {\cal B}- {\cal B}{\cal A}$.
The last step in Eq.~(\ref{LIt}) is easily proved by induction and it is 
known as the 
Hadamard's lemma for exponentials of operators.
In~\cite{bJMP59} $\tilde{\cal L}_I(t)$ of Eq.~(\ref{LIt}) is also called the Lie evolution of the operator ${\cal L}_I$ along the Liouvillian ${\cal L}_a$, for a time $-t$. For further use, we note
 that the Lie evolution of a product of operators is the product of the Lie evolution of the individual  operators:
\begin{equation}
e^{{\cal A}^\times t} [{\cal B}{\cal C}]=e^{{\cal A}^\times t} [{\cal B}]\,e^{{\cal A}^\times t} [{\cal C}]\nonumber .
\end{equation}
%
%Defining $P(X;t):=\langle P_\xi(X;t)\rangle$ and assuming  that at the initial time $t=0$   
%$P(x;0)$ {\em does not depend on the possible values of the process $\xi$} (or that we waited long enough to make 
% the initial conditions irrelevant), Eq.~(\ref{app:liouvilleEq.1}), at the first non vanishing power of $\epsilon$,  
% leads to: 
%%  
%\begin{align}  
%\label{FP_F2_int} 
%\partial_t \tilde{P}(X;t) 
%=  \epsilon^2 \langle \xi^2\rangle \tilde {\cal L}_I(t) \int_0^{t}\mbox{d}u\,  \tilde {\cal L}_I(u) \varphi(t-u){P} (X;t) 
%\end{align}  
%% 
%from which, getting rid of the interaction picture, we have
Integrating Eq.~(\ref{app:liouvilleEq.1}) and averaging over the
realization of $\xi(t)$, we get
\begin{equation}
\label{eq:texp}
\tilde{P}(X;t) = 
\langle \overleftarrow{\exp} \left[ \int_0^t du \; \xi(u) \tilde{\cal L}_I (u) \right]
\rangle P(X;0)
\end{equation}
where $\overleftarrow{\exp} [...]$ is the standard chronological ordered 
exponential (from right to left), $P(X;t) := \langle P_{\xi} (X;t)\rangle$
and we assumed that $P_{\xi} (X;0) = P(X;0)$, i.e. at the initial time
$t=0$ $P_{\xi} (X;0)$ does not depend on the possible values of the
process $\xi$, or that we wait long enough to make the initial conditions 
irrelevant. The r.h.s. of Eq.~(\ref{app:liouvilleEq.1}) can be considered as a 
sort of generalized moment generating function for the fluctuating operator
$\xi(u) \tilde{\cal L}_I$ to which it is possible to associate a generalized
cumulant generating function~\cite{bJMP59}. Keeping up to the
second generalized cumulants leads to the following result~\cite{bJMP59} (note that we use the assumption  $\langle\xi^2\rangle \tau=1$, from which  $\langle\xi^2\rangle=1/ \tau$) 
\begin{equation}
\partial_t \tilde{P}(X;t) = \epsilon^2\, 
\tilde{\cal L}_I(t) \int_0^t du \; \tilde{\cal L}_I(u) \varphi(t-u) P(X;t)
\label{dtp:interaction}
\end{equation}
which coincides with the usual one obtained using a second
order in $\epsilon,$ Zwanzig projection approach~\cite{bJSTAT2015}.

Getting rid of the interaction picture, from Eq.~(\ref{dtp:interaction}) 
we obtain 
\begin{align}  
\label{FP_F2} 
&\partial_t {P}(X;t) \nonumber \\
&= {\cal L}_a P(X;t)+ \epsilon^2 \, \partial_X\,\frac{ 1}{\tau} \int_0^{t}\mbox{d}u\,  e^{{\cal L}_a^\times u}[\partial_X] \varphi(u){P} (X;t).
\end{align}  
This is a standard result, in fact, as we have already stressed, it can be obtained with any perturbation approach, where $\epsilon$ is the small parameter (assuming a finite, but not necessary small, correlation time $\tau$). We
have cited the generalized cumulant approach because it gives a sound
justification of the second order truncation of the full
series of generalized comulants~\cite{bJMP59}). 
The next step is to rewrite, if possible,  Eq.~(\ref{FP_F2}) as the FPE of Eq.(~\ref{FPEGen})
% \begin{align}  
% \label{FP_GEN} 
% \partial_t {P}(X;t) 
% = {\cal L}_a P(X;t)+ \partial_X^2 D(X){P} (X;t).
% \end{align}  
%
\begin{figure}[htb]
        \centering
        \includegraphics[width=.8\linewidth]{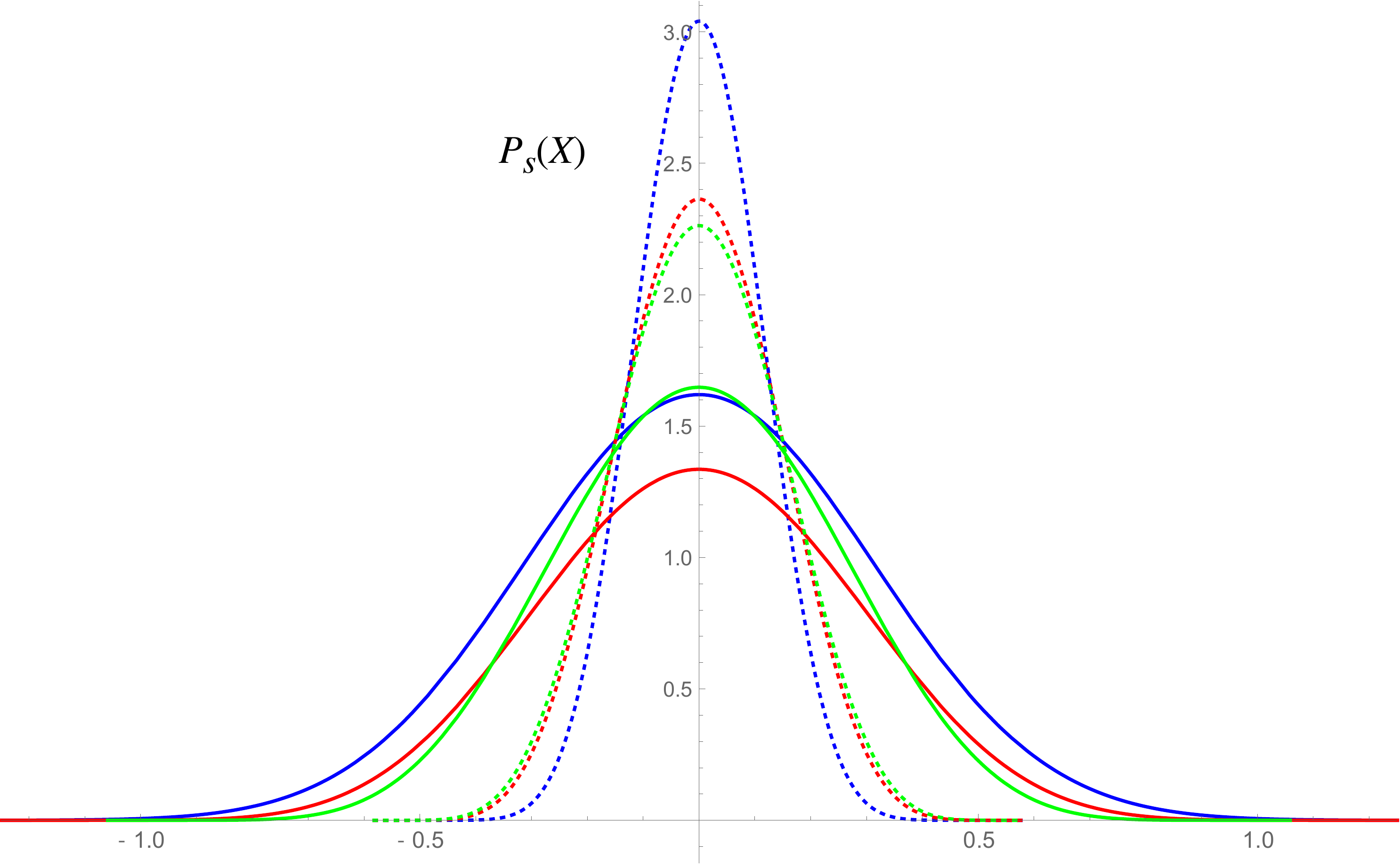}
        \caption{The case where $C(X)=\sinh(X)$, 
                and $\langle \xi^2 \rangle=1$, $\varphi(t)=\exp(-t/\tau)$, $\tau=0.8$ and $\epsilon=0.3$.
                The graphs are the PDFs obtained from Eq.~(\ref{Ps}), in which the  state dependent diffusion coefficient  $D(X)$ is evaluated from 
                Eq.~(\ref{FP_F2}) supplemented 
                with the series expansion of Eq.~(\ref{expLiu_series}) truncated at the fifth order. The solid lines refer to 
                even orders: zeroth (blue), second (red) and fourth (green) one. The dashed lines refer to odd orders: first 
                (blue), third (red) and fifth (green) one.} 
        \label{fig:PDF_VariTau}
\end{figure}
To go from Eq.~(\ref{FP_F2}) to the FPE of Eq.~(\ref{FPEGen}), the crucial term is the operator $e^{{\cal L}_a^\times u}[\partial_X]$.
In most papers using the Zwanzig projection method (e.g.,~\cite{grigo_memory}), the explicit FPE is  obtained from  
Eq.~(\ref{FP_F2}) assuming that $\tau$, identified with the decay time  of the correlation function $\varphi(t)$, is much smaller than
the time scale of
the unperturbed dynamics driven by the Liouvillian ${\cal L}_a$. In this case it is possible to replace,
 in Eq.~(\ref{FP_F2}), the power expansion  (note the shorthand $(\partial_XC(X)):=C'(X)$)
\begin{align}
\label{expLiu_series}
e^{{\cal L}_a^\times u}[\partial_X]&=\partial_X+[{\cal L}_a,\partial_X]\,u + O(u^2) \nonumber \\
&=\partial_X-\partial_X\,C'(X)\,u + O(u^2).
\end{align}
that leads to a FPE with a  state dependent diffusion coefficient,  given by a series of ``moments''
of the time $u$, weighted with the correlation function $\varphi(u)$. 
However, such a series, as it is apparent from Eq.~(\ref{expLiu_series}), contains secular terms and is (generally) not absolutely convergent. This is clearly shown in the 
example considered in Fig.~\ref{fig:PDF_VariTau}. 
A way to avoid this problem is to solve, without approximations,  the Lie evolution of the differential operator $\partial_X$ along the Liovillian ${\cal L}_a$.
In~\cite{bJMP59} this was done for the general case of multidimensional systems and multiplicative forcing. 
In the present simpler one-dimensional case, recalling that ${\cal L}_a=\partial_X C(X)$,
the Lie evolution of ${\partial_X}$, without approximations, 
can be obtained directly as follows:
\begin{align}
\label{Lax2}
&e^{{\cal L}_a^\times u}[\partial_X]=e^{{\cal L}_a^\times u}[\partial_X C(X) \frac{1}{C(X)}  ] \nonumber \\
&= e^{{\cal L}_a^\times u}[{\cal L}_a]\,e^{{\cal L}_a^\times u}[\frac{1}{C(X)} ]
= \partial_X C(X) \frac{1}{C(X_0(X;-u))}
\end{align}
where $X_0(X;-u):=e^{{\cal L}_a^\times u}[X]=\left(e^{-{\cal L}_a^+ u}X\right)$~is the unperturbed backward  evolution, for a time $u$, of the variable of interest, starting from the $X$ position
 at the initial time $u=0$. In the second line of Eq.~(\ref{Lax2}) we have used two trivial facts (see again~\cite{bJMP59} for details and generalizations):
\begin{itemize}
        \item given two operators ${\cal A}$ and  ${\cal B}$,  ${\cal B}$ does not Lie-evolve along ${\cal A}$ when $[{\cal A},{\cal B}]=0$, thus $e^{{\cal L}_a^\times u}[ {\cal L}_a]={\cal L}_a$,
        \item the Lie evolution along a deterministic 
        (first order partial differential operator) Liouvillian of a regular function $C(X)$, is just the back-time evolution of $C(X)$      along the flow generated by the same Liouvillian: 
        \begin{equation}
e^{{\cal L}_a^\times u}[C(X) ]=C(X_0(X;-u)).
        \end{equation}
\end{itemize}
Inserting Eq.~(\ref{Lax2}) in  Eq.~(\ref{FP_F2}) we get, in a clear and straight  way, the BFPE of Lopez, West and Lindenberg~\cite{lwlPRA37} \footnote{actually,
the derivation shown here is a generalization, since we 
do not assume that $\varphi(t)=\exp(-t/\tau)$ and we do not take the limit 
$t \rightarrow \infty$
in the time integration.}:
\begin{align}  
\label{FP_F3} 
&\partial_t {P}(X;t) 
= {\cal L}_a P(X;t) \nonumber \\
&+ \epsilon^2 \, \partial_X^2\,\frac{ 1}{\tau} C(X)\left( \int_0^{t}\mbox{d}u\,\frac{ 1}{C(X_0(X;-u))}\varphi(u) \right) {P} (X;t), 
\end{align}  
namely, the FPE of Eq.~(\ref{FPEGen}) with the state and time dependent 
diffusion coefficient
\begin{equation}
\label{DBFPEt}
D(X,t)_{BFPE}=\epsilon^2 \frac{ 1}{\tau}  C(X)\left( \int_0^{t}\mbox{d}u\,\frac{ 1}{C(X_0(X;-u))}\varphi(u) \right)\\
\end{equation}
that, for large times, becomes
\begin{align}
\label{DBFPE}
D(X,\infty)_{BFPE}=\epsilon^2 \frac{ 1}{\tau}  C(X)\left( \int_0^{\infty}\mbox{d}u\,\frac{ 1}{C(X_0(X;-u))}\varphi(u) \right).
\end{align}
For weak enough noise intensity $\epsilon$, the BFPE looks like 
the best possible 
approximation we can get from a perturbation approach to the SDE of Eq.~(\ref{SDEGen}).
However, this is not the case: the diffusion coefficient in Eq.~(\ref{DBFPE})
turns out to be wrong, as we are going to show. 

\begin{figure*}[htb]
        \centering
        \includegraphics[width=1.0\linewidth]{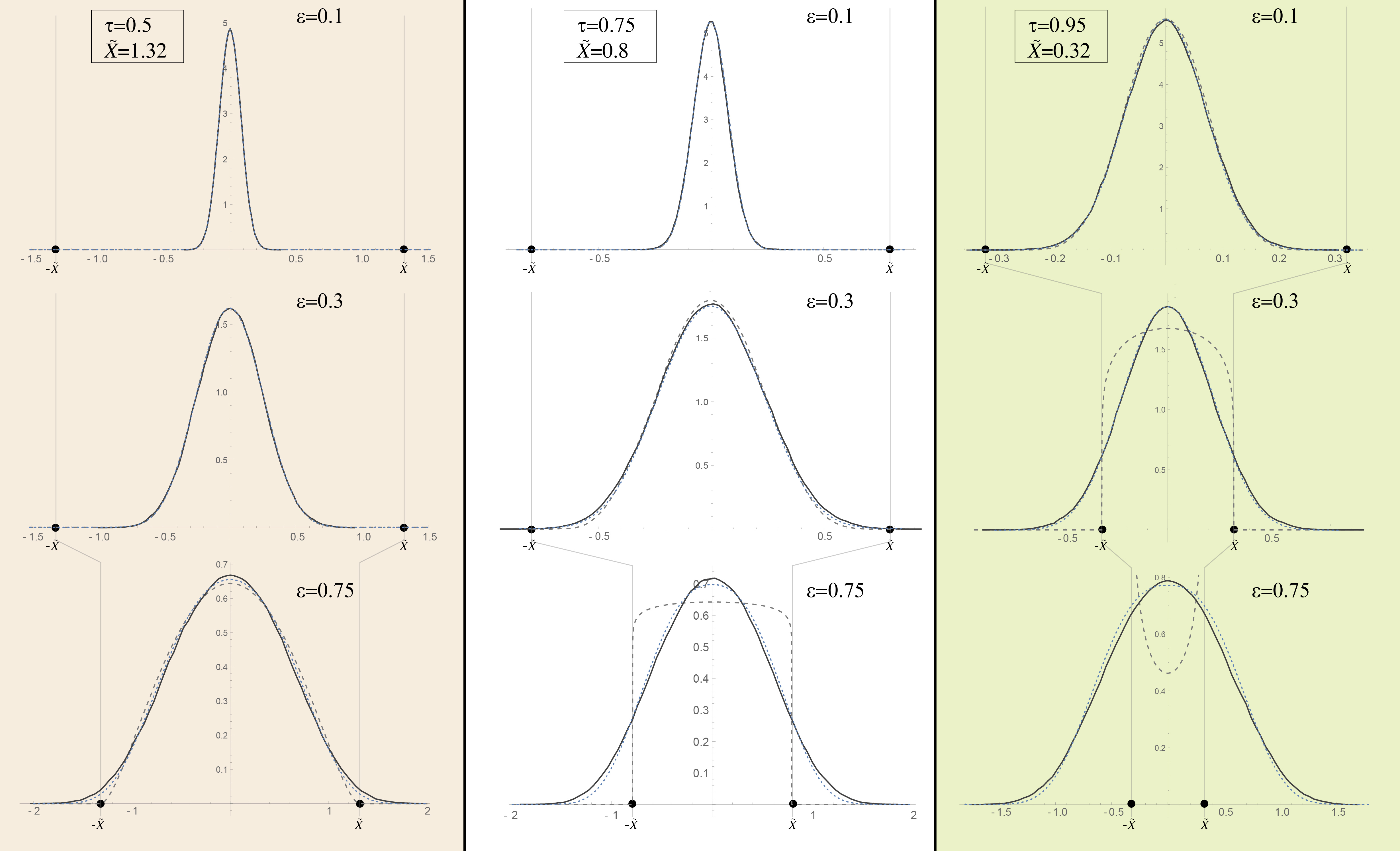}
        \caption{Solid black lines: the stationary PDF from the 
        numerical simulation of the SDE of Eq.~(\ref{SDEGen}) with $C(X)=\alpha
 \sinh(kX)$ and  $\alpha=k=1$. Dashed gray lines: the BFPE stationary PDF 
$P_s(X)_{BFPE}$ from Eq.~(\ref{DBFPESinh}), 
the interval $-\tilde X<X<\tilde X$ is the support of this PDF (see text). Dotted blue 
lines:  
the cBFPE stationary PDF $P_s(X)_{cBFPE}$ (discussed further down in this paper) 
obtained from Eq.~(\ref{Ps}) using 
$D(X)=D(X,\infty)_{cBFPE}$ 
of Eq.~(\ref{DcorBFPESinh}). Note how the BFPE PDF 
 completely fails when, increasing $\epsilon$, the
         width of the PDF becomes comparable (or larger) than the interval width $2\tilde X$, whereas there is an
excellent agreement between simulations and cBFPE PDF for $\tau$
and  $\epsilon$ considered.}
        \label{fig:comparison_cor}
\end{figure*}

It is actually known that in many cases of  interest the diffusion coefficient $D(X)_{BFPE  }$, given in Eq.~(\ref{DBFPE}), becomes negative, giving rise to a non physical negative PDF. A simple example may serve for illustration. Let us consider the case in which $C(X)= \alpha \sinh(kX)$ and  
$\varphi(t)=\exp(-t/\tau)$, with $\alpha>0$. 
The corresponding SDE is related to a well known chemical reaction scheme, 
see~\cite{lwPA119}. A straightforward calculation leads to 
$C(X)/C(X_0(X;-u))=\cosh (\alpha  k u)-\cosh (k X) \sinh (\alpha  k u)$, which
 inserted in Eq.~(\ref{DBFPE}), for times 
$t>>\tau/(1-\alpha k \tau)$, gives ($\theta := \alpha k \tau$)
\begin{equation}
\label{DBFPESinh}
D(X,\infty)_{BFPE}=\epsilon^2 \, \frac{1-\theta  \cosh (k X)}{1-\theta },
\end{equation}
 with the constraint $\theta<1$. From Eq.~(\ref{DBFPESinh}) we see that for $X=\pm\tilde X$, with $\tilde X:=\frac{\log \left(\sqrt{\theta ^2-1}+\theta \right)}{k}$, 
 the diffusion coefficient of the BFPE vanishes and
 for $|X|>\tilde X$ it is negative 
which is clearly unphysical. 
Using Eq.~(\ref{DBFPESinh}) in Eq.~(~\ref{Ps}), we obtain the stationary PDF: 
\begin{equation}
\label{PsSinh}
P_s(X)_{BFPE}=\frac{1}{Z_{BFPE}}\left(\frac{1-\theta  \cosh (k X)}{1-\theta }\right)^{\frac{1-\theta^2-k^2 \tau ^2 \epsilon ^2}{k^2  \tau\, \epsilon ^2}}
\end{equation}
that is affected by the same problem for $|X|>\tilde X$. The standard way 
to cure this flaw of the BFPE is to restrict the support of the
 PDF~\cite{lwPA119,lwlPRA37}.
In this case, for example, 
the first and the second derivatives of Eq.~(\ref{PsSinh}) 
 vanish in $|X|=\tilde X$, therefore one could limit the support of the 
PDF of Eq.~(\ref{PsSinh}) to $X\in (-\tilde X, \tilde X)$. However, 
from Fig.~\ref{fig:comparison_cor} it is clear
that by increasing $\epsilon$, the result of Eq.~(\ref{PsSinh}) does not agree well with that obtained from 
the numerical simulation
of the SDE of Eq.~(\ref{SDEGen}). Only for very small values of $\tau \epsilon$ the result is good (i.e, when
the width of the PDF is small compared to $2 \tilde X$).
The same problem is present when other drift fields $C(X)$ are considered:
the case of $C(X)= X^3$ is shown in Appendix~\ref{app:cubic}, other examples 
can be found in the literature~\cite{mwlPRA35,tgPRL61,tgPRA38,glmmmp38}). 

\section{The cured BFPE}\label{sec:curedBFPE}

We show in this section that the flaws of
the BFPE are due to an incorrect implementation of the perturbation procedure, and we will cure this situation.  

Note first that the possibly negative $D_{BFPE}$ value of Eq.~(\ref{DBFPE}) is due to the fact that the kernel 
of the integral can be negative for some $X$ values.
\begin{figure}[hbt]
        \centering
        \includegraphics[width=0.7\linewidth]{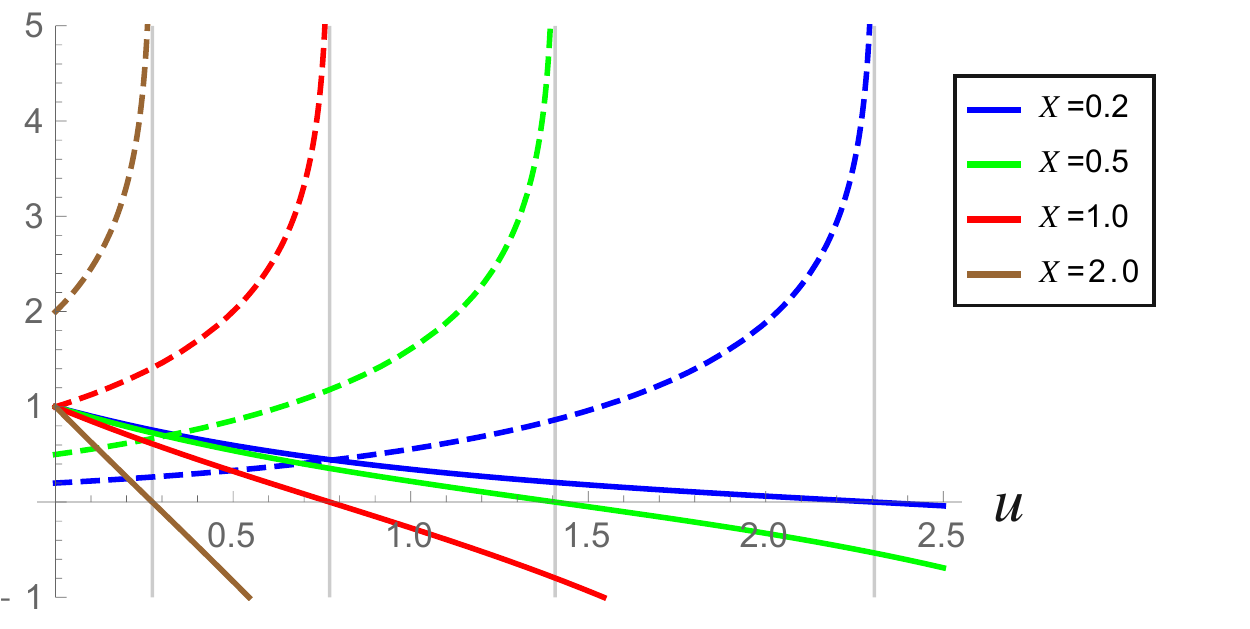}
        \caption{Case  $C(X)=\sinh(X)$.
                Solid colored lines: the function $C(X)/C(X_0(X;-u))=\cosh (u)-\sinh (u) \cosh (X)$  for different initial positions $X_0(X;0)=X$.
                Dashed colored lines: the back time evolution $X_0(X;-u)=2 \coth ^{-1}\left(e^{-  u} \coth \left(\frac{X}{2}\right)\right)$, 
                for the same different initial values $X_0(X;0)=X$.   
                Thin gray vertical lines:  asymptotes at the corresponding time values 
                $\bar{u}(X)=\log \left(\sqrt{\frac{\cosh (X)+1}{\cosh (X)-1}}\right)$. 
                At the time value $\bar{u}(X)$ where the back time evolution $X_0(X;-u)$ diverges, the function $C(X)/C(X_0(X;-u))$ vanishes. For larger times it is a negative number.}
        \label{fig_c}
\end{figure}

\begin{figure}[hbt]
        \centering
        \includegraphics[width=0.7\linewidth]{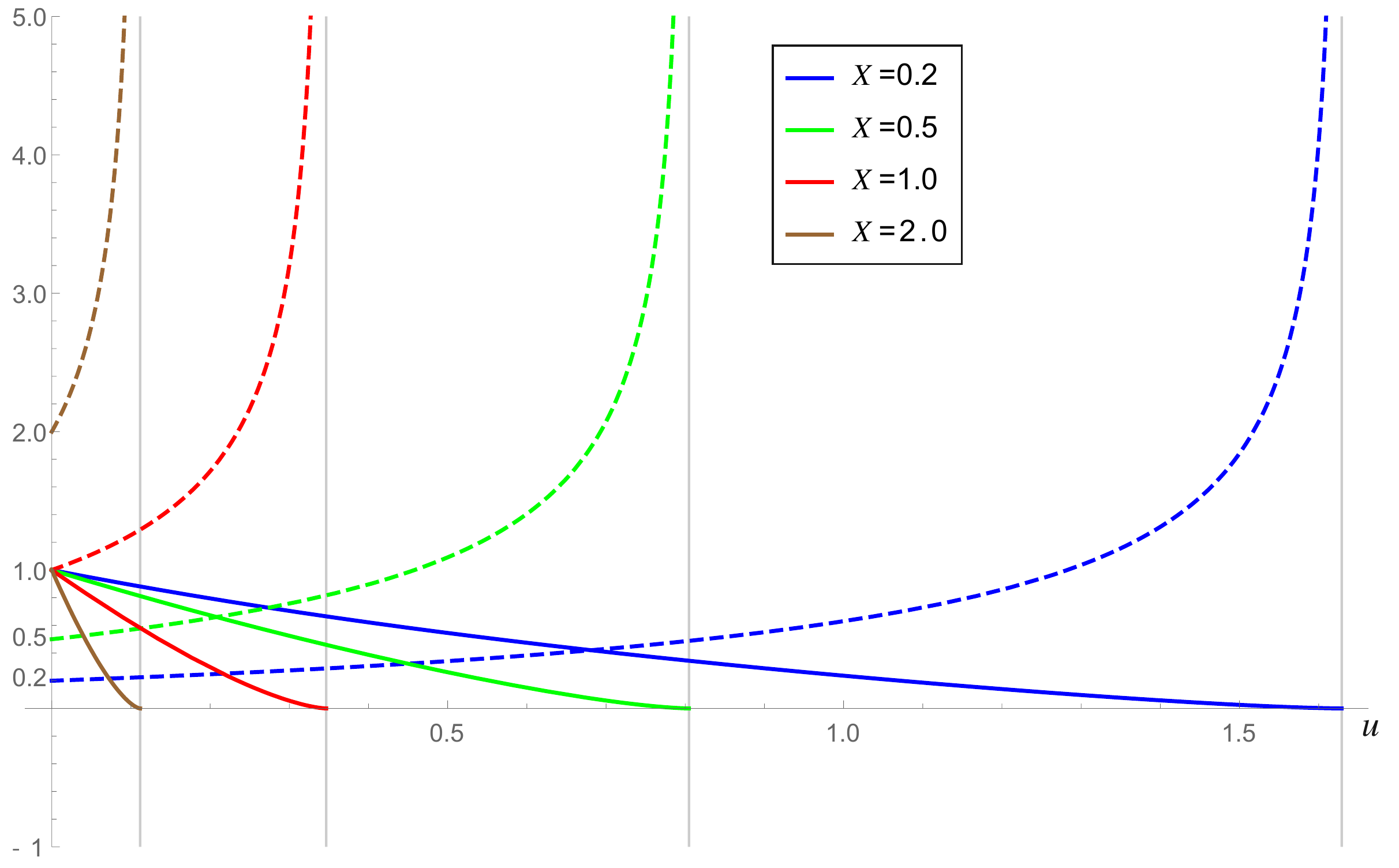}
        \caption{Case $C(X)= \alpha X^3$ with $\alpha=1$.
                Solid colored lines: the function  $C(X)/C(X_0(X;-u))=e^{u/2} \left( -2   X^2 \sinh (u)-\sinh (u)+\cosh (u)\right)^{3/2}$  for different initial positions $X_0(X;0)=X$.
                Dashed colored lines: the back time evolution $X_0(X;-u)=X/\sqrt{-X^2  + e^{-2 u} (1 + X^2 )}$, 
                for the same initial values $X_0(X;0)=X$.   
                Thin gray vertical lines:  asymptotes at the corresponding time values 
        $\bar{u}(X)=-\frac{1}{2} \log \left(\frac{  X^2}{ X^2+1}\right)$. 
        At the times $\bar{u}$ when the back time evolution $X_0(X;-u)$ diverges, the function $C(X)/C(X_0(X;-u))$ vanishes, while for larger times it is a complex number.}
        \label{fig:Plot_Cubica_}
\end{figure}

\begin{figure}[htb]
        \centering
        \includegraphics[width=\linewidth]{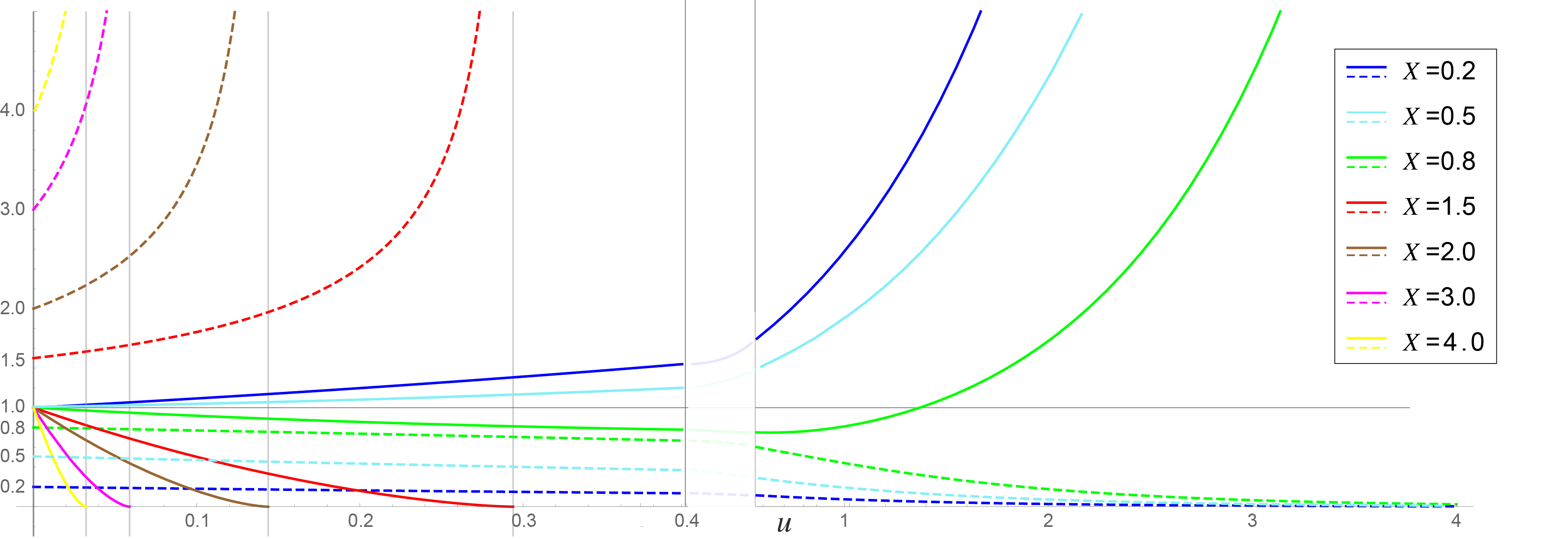}
        \caption{Case $C(X)=X + \alpha X^3$, $\alpha=1$.
                Solid colored lines: $C(X)/C(X_0(X;-u))=e^{-u} (\sqrt{\alpha  \left(e^{2 u}-1\right) X^2+1}\,)^{3}$
                for different initial positions $X_0(X;0)=X$.
                Dashed colored lines: $X_0(X;-u)=e^u X/\sqrt{\alpha  \left(e^{2 u}-1\right) X^2+1}$ for the same initial values $X_0(X;0)=X$.   
                Thin gray vertical lines:  asymptotes at the corresponding time values 
                $\bar{u}=\frac{1}{2} \log \left(\frac{\alpha  X^2-1.}{\alpha  X^2}\right)$.   For $|X|<1$ the backward
                trajectories do not diverge at all and the function $C(X)/C(X_0(X;-u)$ is always positive.}
        \label{fig:XmenoX3}
\end{figure}

\begin{figure}[htb]
        \centering
        \includegraphics[width=0.8\linewidth]{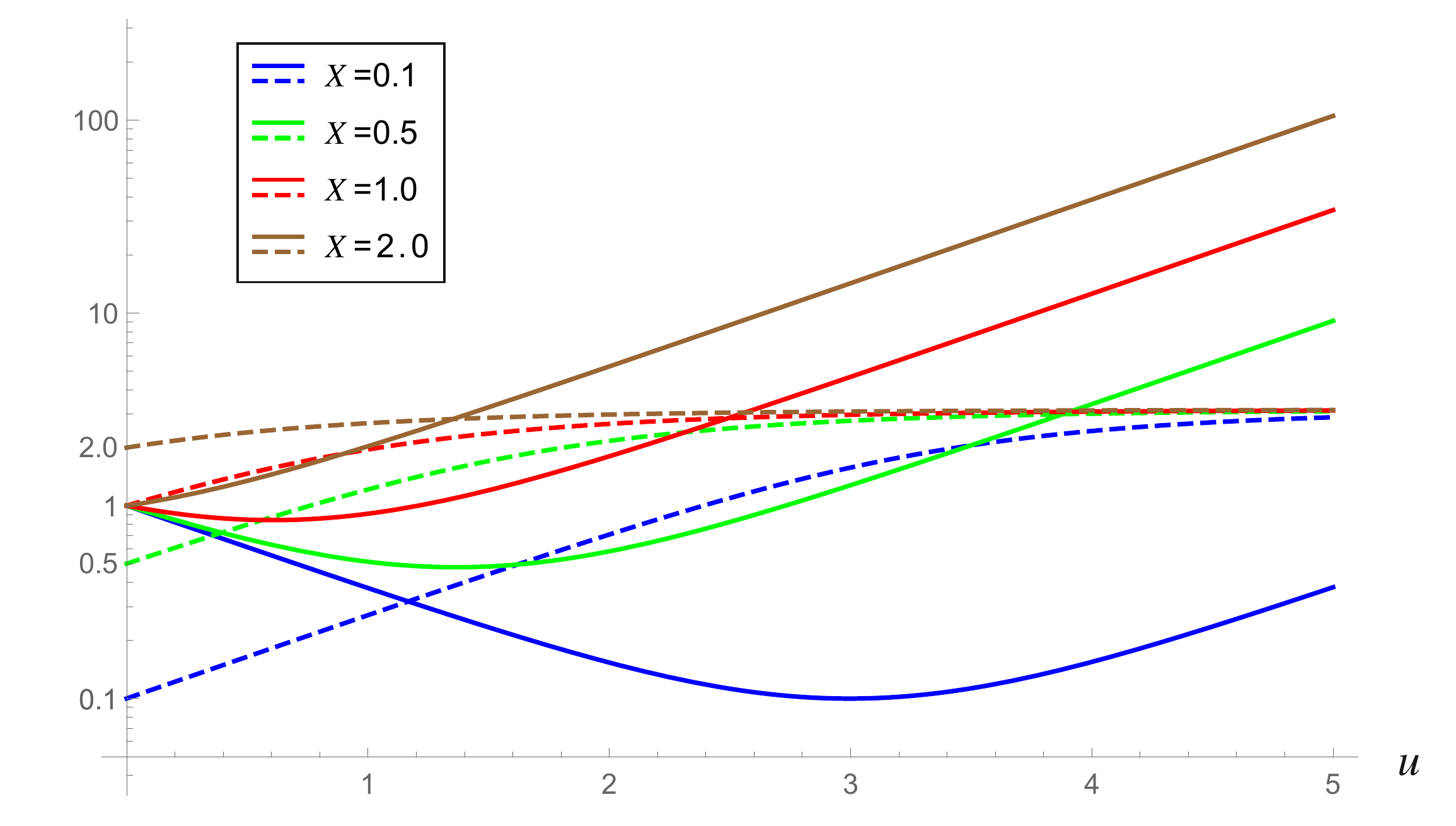}
        \caption{Semi-log plot for the case 
                $C(X)=  \alpha \sin(kX)$, $\alpha=k=1$.
                Solid colored lines: $C(X)/C(X_0(X;-u))=\cosh(k \alpha u)-\sinh (k\alpha u) \cos(kX)$  for different initial positions $X_0(X;0)=X$.
        Dashed colored lines: the back time evolution $X_0(X;-u)=2 \cot^{-1}\left(e^{- k\alpha u}
         \cot \left(\frac{kX}{2}\right)\right)/k$, 
        for the same initial values $X_0(X;0)=X$.   }
        \label{fig:LogPlot_sinx}
\end{figure}

Considering once more the case of $C(X)=\alpha \sinh(k X)$, we see from Fig.~\ref{fig_c}, solid lines, that, after a given time 
$\bar{u}(X)$, the function $C(X)/C(X_0(X;-u))$ becomes negative. Note also that the larger the $X$ value, the shorter 
the time $\bar{u}(X)$. Thus, whatever the correlation decay time $\tau\in (0,1/\alpha k)$, there will always be a certain $\tilde{X}$ 
value such that $D(X,\infty)_{BFPE}$ of Eq.~(\ref{DBFPE})  is negative for 
$|X|>\tilde{X}$ (the greater the $\tau$ value, the smaller the $\tilde X$ value).

%
%What happens in general may be rather complicated, with very different scenarios
%depending on $C(X)$.

Depending on $C(X)$, we may have rather different scenario:
for example, when $C(X)= X^3$ for $|X|>\tilde X$,  the  kernel of the $D(X,\infty)_{BFPE}$ of Eq.~(\ref{DBFPE}), turns out to be a 
 complex number see Appendix~\ref{app:cubic} and Fig.~\ref{fig:Plot_Cubica_}. Therefore, in this case it would 
 seem that the BFPE does not exist at all. 

Other interesting examples are the case when $C(X)=X+ \alpha X^3$ (see
 Fig.~\ref{fig:XmenoX3}), where, depending on
the initial $X$, the kernel can go negative ($|X>1|$) or stay positive ($|X<1|$); and the case when $C(X) = \alpha \sin(kX)$, where the kernel is always 
positive (see Fig.~\ref{fig:LogPlot_sinx}).

The  shortcomings of the BFPE are however artifacts, introduced
by an unapropriate use of the interaction picture, and they can be fixed.

%bistable systems.
When we go to the interaction picture and then return to the normal
representation, we time evolve the variable of interest forth and back, 
along the flow generated by the $-C(X)$ drift field. 

The backward evolution is indicated by $X_0 = X_0 (X;-u)$. Using 
Eq.~(\ref{SDEGen}) we can easily invert this relation, to get
$u(X,X_0) = \int_X^{X_0} \frac{1}{c(y)} \; dy$. We define the
$X$ dependent time $\bar u(X)$ as the time it takes the
unperturbed evolution, starting from $X$, to go to $X_0 \rightarrow \infty$,
namely $\bar u(X) := u(X,\infty) = \int_{X}^{\infty} \frac{1}{c(y)} \; dy$.
For a dissipative flow asymptotically linear, namely with
$\lim_{X \rightarrow \infty} C(X) \propto X$, $u(X)$ is clearly infinite:
starting from any position $X$, it takes an infinite time to go
backward to $X_0 \rightarrow \infty$. However, if 
$\lim_{X \rightarrow \infty} C(X) > X^h$, $h \ge 1$, 
we have a finite value for $\bar u(X)$: 
going back in time,  the trajectory $X_0(X;-u)$ 
in a finite time $\bar{u}(X)$ reaches all possible values, greater than $X$.
%For a dissipative flow, with an absolute local divergence that increases asymptotically  along the same flow, the  
%backward evolution $X_0(X;-u)$, starting
%from the initial position $X_0(X;0)=X$, diverges at some $X$ dependent time $\bar{u}(X)$. Thus, going back in time, in a finite time $\bar{u}(X)$
% the trajectory $X_0(X;-u)$    
For example, in the case where $C(X)=\alpha \sinh(kX)$
 we show in Fig.~\ref{fig_c}, dashed lines, that   
$X_0(X;-u)=\frac{2}{k} \coth ^{-1}\left(e^{-\alpha  k u} \coth \left(\frac{k X}{2}\right)\right)$ 
has an asymptote at 
$u=\bar{u}(X):=\frac{1}{k \alpha}\log \left(\sqrt{\frac{\cosh (kX)+1}{\cosh (kX)-1}}\right)$ 
(the case $C(X)=X^3$ is shown in Fig.~\ref{fig:Plot_Cubica_}, and
the case $C(X)= X + \alpha X^3$ in Fig.~\ref{fig:XmenoX3}). 
For ``preceding'' times $-u$ with $u>\bar{u}(X)$ there are no points in 
the state-space that are 
connected to $X$ by the flow generated by the drift field $-C(X)$. 
This is obviously due to the strong irreversible nature of the  flow, that shrinks 
the state space.
In essence, this implies that for such strongly dissipative flows, the backward evolution must be limited to times
$u<\bar{u}(X)$, i.e. \textit{we must multiply  any function of $X_0(X;-u)$
by the Heaviside function $\Theta(\bar{u}(X)-u)$. }
\begin{widetext}
         Therefore, 
 the BFPE state dependent diffusion coefficient of Eqs.~(\ref{DBFPEt})-(\ref{DBFPE}) must be corrected  as follows (cBFPE stands for corrected BFPE):
\begin{align}
\label{Dcort}
D(X,t)_{cBFPE}&=
\epsilon^2 \frac{ 1}{\tau}C(X)\left( \int_0^{t}\mbox{d}u\,
\frac{ \Theta(\bar{u}(X)-u)}{C(X_0(X;-u))}\varphi(u) \right)
\end{align}
\begin{align}
\label{Dcor}
D(X,\infty)_{cBFPE}&= 
\epsilon^2 \frac{ 1}{\tau}  C(X)\left( \int_0^{\bar{u}(X)}\mbox{d}u\,
\frac{ 1}{C(X_0(X;-u))}\varphi(u) \right)\nonumber \\
&=D(X,\bar{u}(X))_{BFPE}
\end{align}
\textit{Eqs.~(\ref{Dcort})-(\ref{Dcor}) are the main result of the present work}.
Concerning the stationary PDF, the correct result is obtained using Eq.~(\ref{Dcor}) in Eq.~(\ref{Ps}).

For the  case  $C(X)=\alpha \sinh(kX)$, 
from Eq.~(\ref{Dcor}) we get :
\begin{align}
\label{DcorBFPESinh}
&D(X,\infty)_{cBFPE}= \nonumber \\
&\epsilon^2 \frac{1}{1-(\alpha k\tau) ^2}  \left(\alpha k\tau  (\cosh (kX)+1) \left| \tanh \left(\frac{kX}{2}\right)\right| ^{\frac{\alpha k\tau +1}{\alpha k\tau }}-\tau  \cosh (kX)+1\right).
\end{align}
\end{widetext}
The  state dependent diffusion coefficient $D(X,\infty)_{cBFPE}$ of Eq.~(\ref{DcorBFPESinh}) is always positive.
The stationary PDF for this case is obtained using Eq.~(\ref{DcorBFPESinh}) in Eq.~(\ref{Ps}). Because of the
integral in the exponent  in Eq.~(\ref{Ps}), an analytical expression cannot be obtained: the results of numerical integration, for different values of $\tau$ and $\epsilon$, are shown in Fig.~(\ref{fig:comparison_cor}). 
We can see that the stationary PDFs of the corrected BFPE are quite close 
to those obtained from the numerical integration of the SDE, even 
for large $\tau$ values and relatively large $\epsilon$.
%
%\begin{figure*}[htb]
%        \centering
%        \includegraphics[width=1.0\linewidth]{sinh_varitau_epsilon_cor}
%        \caption{The same of Fig.~\ref{fig:comparison}  but  the dashed blue lines now refer to the corrected
%                BFPE stationary PDF ($P_s(X)_{BFPE}^{cor}$) obtained from Eq.~(\ref{Ps}) using $D(X)=D(X,\infty)_{BFPE}^{cor}$ 
%                of Eq.~(\ref{DcorBFPESinh}). The dashed light gray lines correspond to the $P_s(X)_{BFPE}$ curves 
%                from Eq.~(\ref{DBFPESinh}). Note that the $P_s(X)_{BFPE}^{cor}$ is 
%well behaved even for $\alpha k \tau>1$.}
%        \label{fig:comparison_cor}
%\end{figure*}
%
In the case of
$C(X)= X^3$, $D(X,\infty)_{cBFPE}$ of Eq.~(\ref{Dcor}) and the corresponding stationary PDF are now real quantities, see Appendix~\ref{app:cubic} and 
Figs.~\ref{fig:Dcubic} and~\ref{fig:cubica_comparison_cor}.

\begin{figure}[htb]
        \centering
        \includegraphics[width=0.7\linewidth]{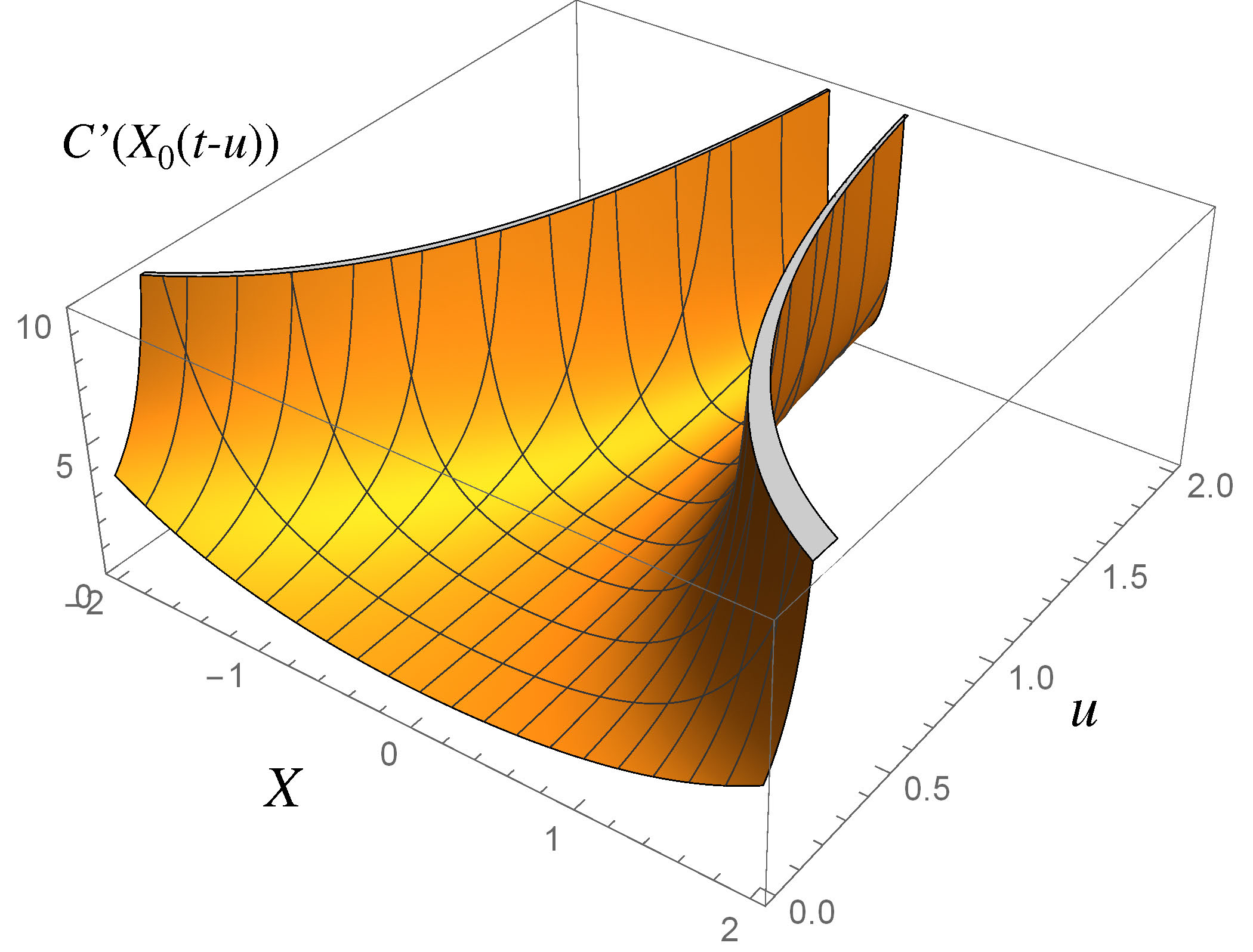}
        \caption{$C'(X)$ along the backward evolution
                $X_0(X;-u)$, $X_0(X;0)=X$, for the hyperbolic drift fields $C(X)=\alpha \sinh(k X)$, $\alpha=k=1$. }
        \label{fig:divSinh}
\end{figure}
\begin{figure}[htb]
        \centering
        \includegraphics[width=0.7\linewidth]{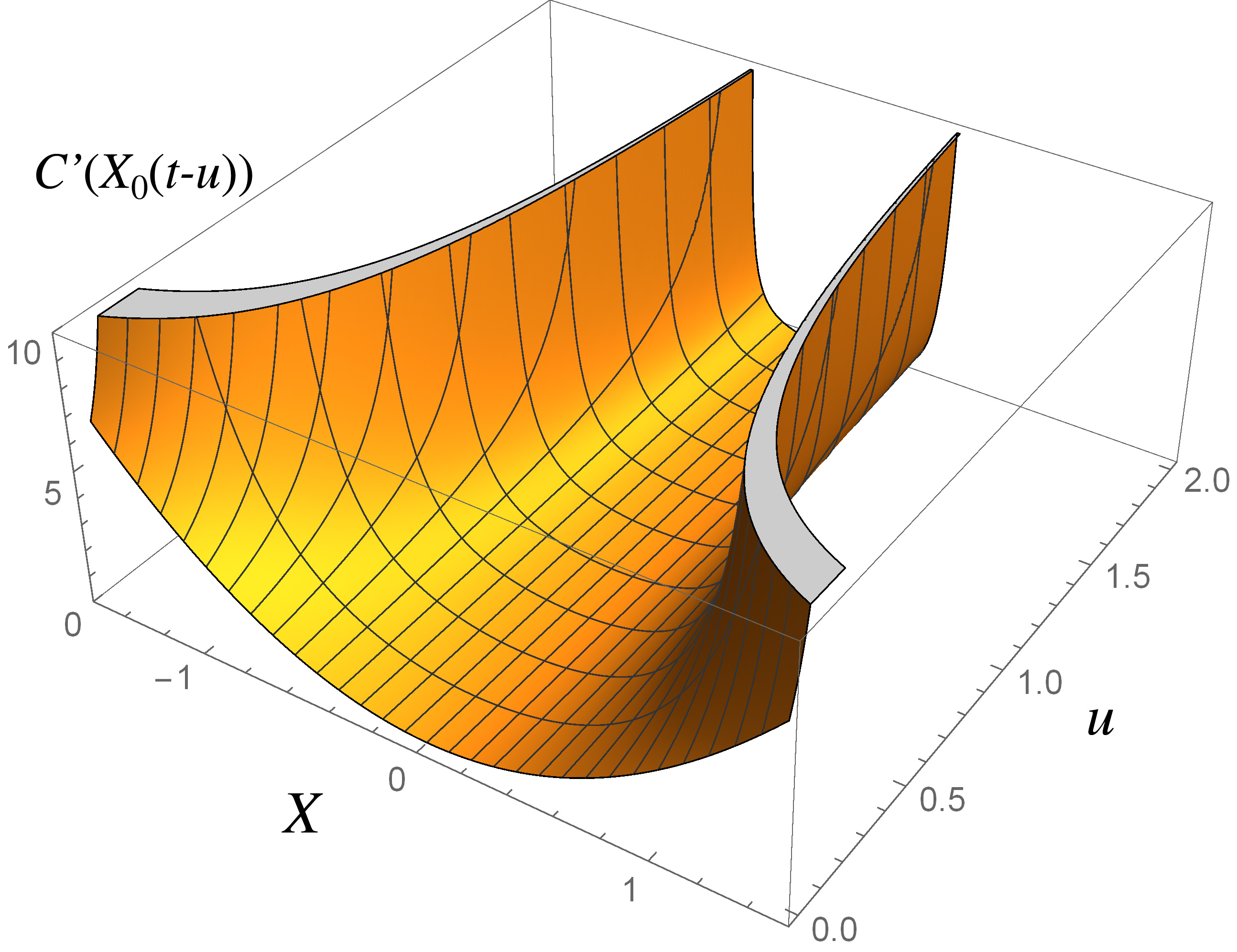}
        \caption{$C'(X)$ along the backward evolution
                $X_0(X;-u)$, $X_0(X;0)=X$, for the pure cubic drift fields $C(X)=\alpha X^3$,$\alpha=1$. }
        \label{fig:divmenoX3}
\end{figure}

\begin{figure}[htb]
        \centering
        \includegraphics[width=0.7\linewidth]{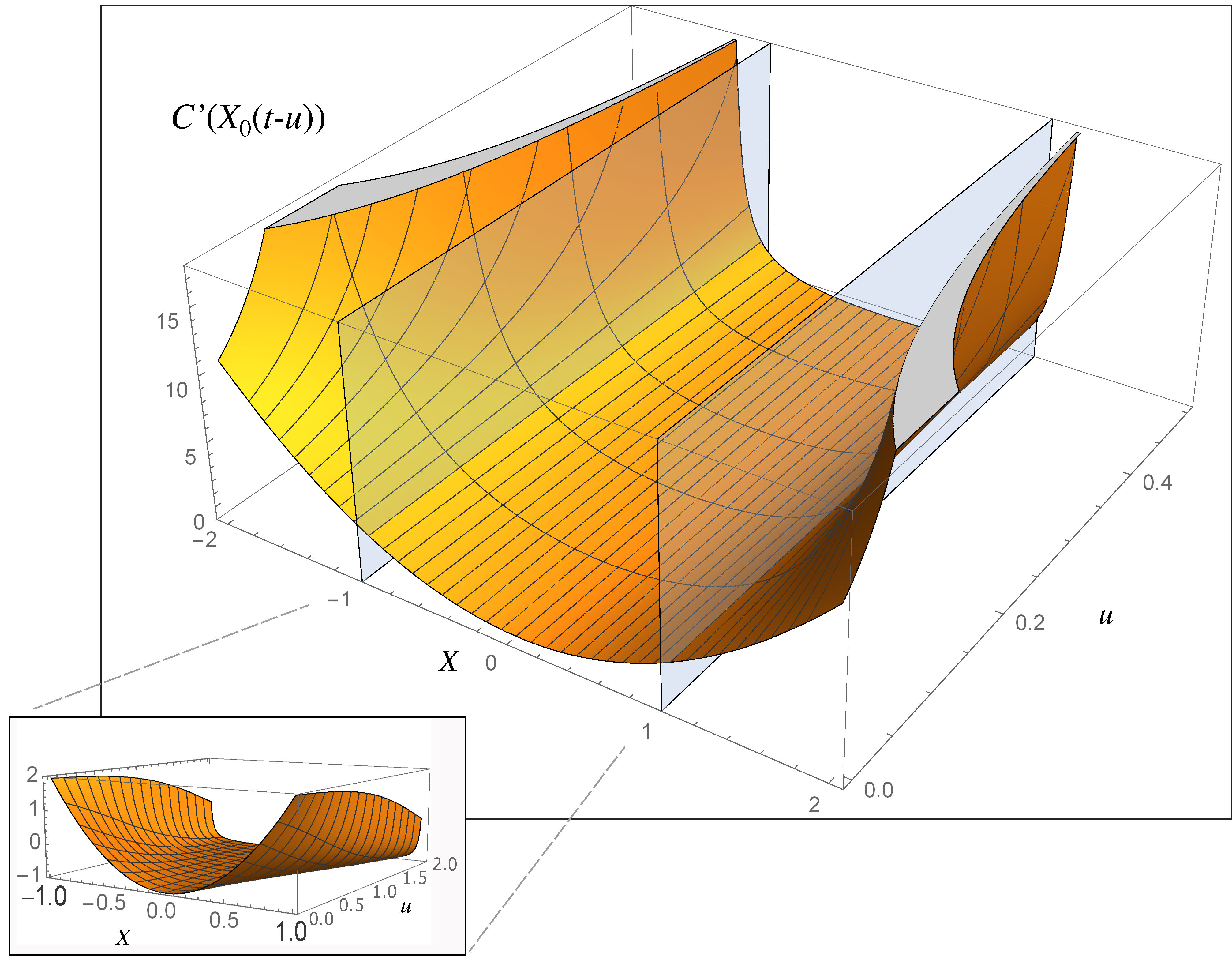}
        \caption{$C'(X)$ along the backward evolution
                $X_0(X;-u)$, $X_0(X;0)=X$, for the drift fields $C(X)=X+\alpha X^3$,$\alpha=1$.  For $|X|<1$ $C'(X_0(X;-u))$ decreases with $u$ (the inset is a magnification of the interval $|X|<1$), while for $|X|>1$ it increases.}
        \label{fig:divXmenoX3}
\end{figure}

We would like to add a few comments about the divergence of 
the backward evolution $X_0(X;-u)$:
we have seen that there are drift fields such that for any initial position
 $X_0(X;0)=X$, the backward evolution diverges with an asymptote at a given finite 
time $\bar u(X)= \int_{X}^{\infty} \frac{1}{c(y)} \; dy$. This behaviour is shown in Figs.~\ref{fig:divSinh},
 \ref{fig:divmenoX3} and \ref{fig:divXmenoX3} for three different drift fields, respectively: in the first two cases 
we observe a divergency for any initial $X$ (that means a finite $\bar u(X)$ value), in the last case there is clearly a
range of $X$ where the backward evolution does not diverge (thus, $\bar u(X)=\infty$). On the other hand,
the important case of Brownian motion in a
periodic potential,  a heuristic model   with  applications in various
branches of science and technology, like the diffusive
dynamics of atoms and molecules on crystal surfaces~\cite{afyAP51}, modelled using
 $C(X)=\alpha \sin(k X)$, is such that $\bar u(X)=\infty$ $\forall X$. In fact,
 the function $C(X)/C(X_0(X;-u))$  is 
always positive and simply increases with $u$  as $e^{k\alpha u}$. 
Therefore in this case the ``standard'' BFPE formula 
of Eq.~(\ref{DBFPEt}) for the diffusion coefficient is correct. 

\section{A comparison with the Local Linearization Approach\label{sec:LLA}}

\begin{figure*}[hbt]
        \centering
        \includegraphics[width=1.0\linewidth]{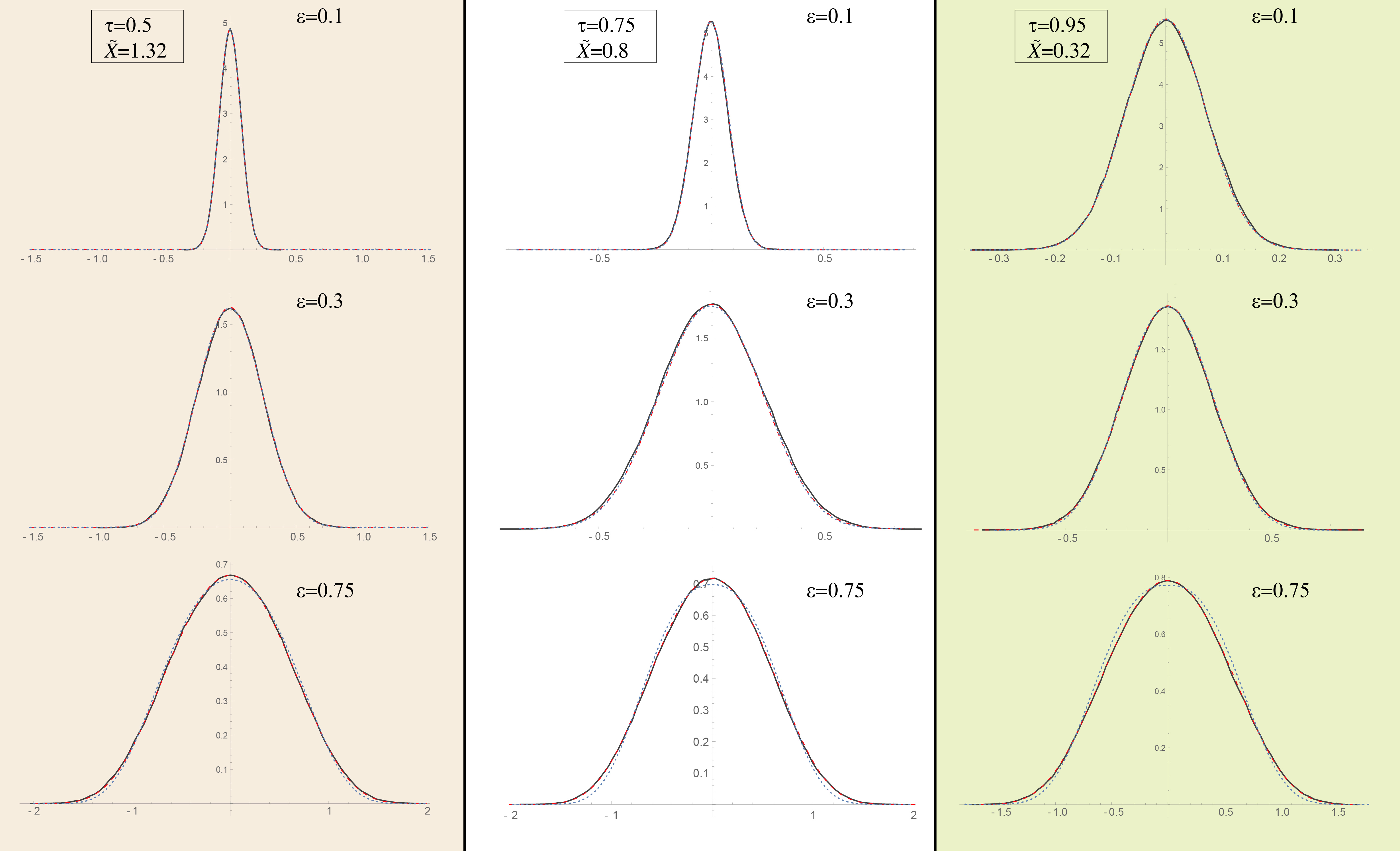}
        \caption{The same as Fig.~\ref{fig:comparison_cor} but without the  
$P_s(X)_{BFPE}$ and with inserted the $P_s(X)_{LLA}$. 
Namely, solid black lines: the stationary PDF from the 
        numerical simulations of the SDE of Eq.~(\ref{SDEGen}) with $C(X)=\sinh(X)$. Dotted blue 
lines:  
the cBFPE stationary PDF $P_s(X)_{cBFPE}$ obtained from Eq.~(\ref{Ps}) using 
$D(X)=D(X,\infty)_{cBFPE}$ 
of Eq.~(\ref{DcorBFPESinh}). Dashed red lines (barely visible close or under the solid lines): $P_s(X)_{LLA}$  of Eq.~(\ref{PsSinhLLA}). The three columns correspond to three 
        different values for $\tau$, while the three rows corresponds to three different values for $\epsilon$. 
        Note the excellent agreement between simulations and LLA PDF.}
        \label{fig:comparison}
\end{figure*}

As we mentioned in the Introduction, very often the LLA FPE turns out to be fairly close to the numerical simulations. This is shown in Fig.~\ref{fig:comparison}, for
the case $C(X)=\alpha \sinh(kX)$. We are going to show that this is not 
a coincidence: as a matter of fact, the LLA FPE is an excellent approximation
of the cBFPE, when the latter is applicable (i.e., typically, small $\epsilon$ and finite, but not small, $\tau$).

We need to briefly go through the derivation of the LLA FPE.
West at al. have shown~\cite{lwlPRA37} that the LLA FPE can be formally derived from the BFPE of  Eq.~(\ref{FP_F3}) as follows:
\begin{enumerate}[label=\alph*]
        \item \label{en:1} there is a large enough time-scale separation
        between the unperturbed dynamics and the decay time of the correlation function $\varphi(t)$, so that
        the unperturbed dynamics $X_0(X;-u)$ can be considered close to the initial position $X$;
        \item \label{en:2} assuming \ref{en:1} above, rather than expanding   $\frac{ 1}{C(X_0(X;-u))}$ in powers of
        $u$ (which would give rise to the same secular terms as the expansion in Eq.~(\ref{expLiu_series})), expand its logarithm
        \begin{align}
        \label{SeriesLog}
        &\frac{ 1}{C(X_0(X;-u))}= e^{\log\left(\frac{ 1}{C(X_0(X;-u))}\right)} \nonumber \\
        &=e^{ \log\left(\frac{ 1}{C(X)}\right)- C'(X)\,u 
                -\frac{1}{2} C(X) C''(X)\,u^2 +O(u^3)},
        \end{align}
        and truncate the series at the first order.
\end{enumerate}
Using  point~\ref{en:2} in Eq.~(\ref{FP_F3}), we are led to the LLA FPE 
(here generalized to finite times and to a generic correlation function of the noise):
\begin{align}  
\label{FP_F4} 
&\partial_t {P}(X;t) 
\sim  {\cal L}_a P(X;t) \nonumber \\
&+ \epsilon^2\frac{1}{\tau}\partial_X^2 \left( \int_0^{t}\mbox{d}u\,
e^{- C'(X) u}\varphi(u) \right) {P} (X;t).
\end{align}  

Note that for $C(X)=\gamma X$, the series expansion of the r.h.s. of 
Eq.~(\ref{SeriesLog}) stops \textit{exactly} at
the first order in $u$, while this does not happen expanding  the term $1/C(X_0(X;-u))$. 
Therefore, instead of using the West and al.
approach (given by \ref{en:1}-\ref{en:2} above) to go from the BFPE to the LLA FPE,  the latter can be 
directly obtained by replacing the function $C(X)/C(X_0(X;-u))$ with an exponential function 
 with state dependent decay coefficient $C'(X)$:
$C(X)/C(X_0(X;-u))\to \exp(-C'(X)u)$).
From Eq.~(\ref{FP_F4}) we get  the following result for the  state dependent diffusion coefficient  of the FPE:
\begin{equation}
\label{DLLAt}
D(X,t)_{LLA}=\epsilon^2 \frac{1}{\tau}  \, \left( \int_0^{t}\mbox{d}u\,
e^{- C'(X) u}\varphi(u) \right)
\end{equation}
that, for large times becomes
\begin{align}
\label{DLLA}
D(X,\infty)_{LLA}=\epsilon^2\, \frac{\hat\varphi(C'(X))}{\tau} 
\end{align}
where $\hat\varphi$ stands for Laplace transform of $\varphi$. From 
Eq.~(\ref{DLLAt})  it turns out that $D(X,\infty)_{LLA}$ exists and 
 is positive under fairly general
conditions. For example, considering again the case $C(X)= \alpha \sinh(kX)$,
from Eq.~(\ref{DLLAt}) we easily get
\begin{equation}
\label{DLLASinh}
D(X,\infty)_{LLA}=\frac{\epsilon^2}{1+\alpha  k \tau  \cosh (k X) },
\end{equation}
where the only constraint is that the flow is not divergent (namely, $\alpha>0$). Using Eq.~(\ref{DLLASinh}) in Eq.~(\ref{Ps}) we obtain the LLA stationary PDF for this case:
\begin{align}
\label{PsSinhLLA}  
&P_s(X)_{LLA}\nonumber \\&=\frac{1}{Z_{LLA}} \left(\frac{1+\alpha  k \tau  \cosh (k X) }{1+\alpha  k \tau }\right) e^{-\frac{\alpha  \sinh ^2\left(\frac{k X}{2}\right) (\alpha  k \tau +\alpha  k \tau  \cosh (k X)+2)}{k \epsilon ^2}},
\end{align}
In Appendix~\ref{app:cubic} we report the LLA results for the cubic case. In Fig.~\ref{fig:comparison} 
we can see the stationary PDFs of the LLA FPE, together with the results from the cBFPE: the agreement with the numerical integration of the 
SDE of Eq.~(\ref{SDEGen}) is very good.

\begin{figure*}[htb]
        \includegraphics[width=1.0\linewidth]{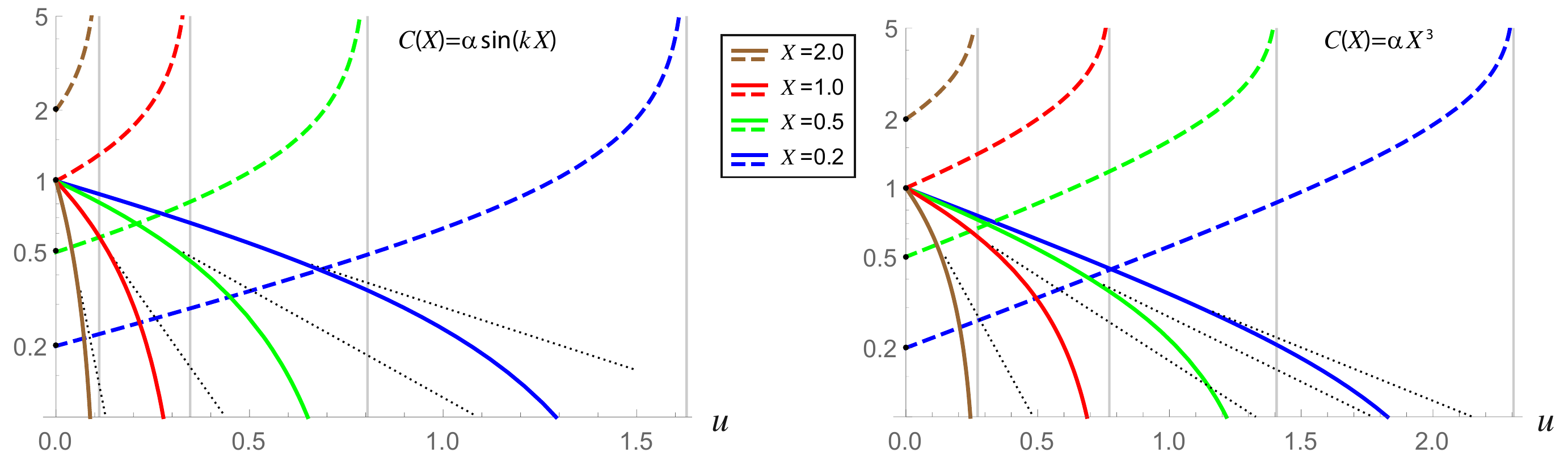}
        \centering
        \caption{Left (right), the same of Fig.~\ref{fig_c} (Fig.~\ref{fig:Plot_Cubica_}) but in log scale. The dotted black lines 
                correspond to the LLA approximation. We can see that the deviation from the exponential decay of the function 
                $C(X)/C(X_0(X;-u))$ (solid lines) is relevant only in the final part, where the value of the function is  
                relatively small.}
        \label{fig:log}
\end{figure*}

Fig.~\ref{fig:log} compares the kernels of the cBFPE and of the LLA for the
cases $C(X)=\alpha\sinh(kX)$ and $C(X)=\alpha X^3$. It turns out that the
LLA kernel (dotted lines) is an excellent approximation of the cBFPE kernel. It is 
hence not surprising that the LLA PDF is as close to the simulations as it 
is the cBFPE PDF.

This is a nice explanation of what has been down heuristically in the literature: 
the LLA approach of Grigolini~\cite{tgPRA38,ffgmJSP52a} is indeed based on the assumption that, for any value of $X$, we can safely replace the 
unperturbed {\em backward} evolution of the function $f(X,u):=C(X)/C(X_0(X;-u))$, with an exponential function of the time $u$, with the $X$ dependent exponent:
$f(X,u)\sim \exp[-C'(X)u]$.
For one-dimensional  dissipative systems, the exponential behavior of such a 
{\em back} time evolution is typical. 

Actually, there is another general argument, not related to the cBFPE, that leads us to speculate that typically (but not always), the LLA FPE  works well, also for strong perturbations. In fact  it is possible to
prove that the LLA  and the Fox 
 functional-calculus~\cite{fPRA33,fPRA34} corresponds to the Almost Gaussian Assumption for generalized stochastic operators~\cite{bJMP59}:
 independently of the value of $\epsilon$, \textit{when $\xi(t)$ is a Gaussian stochastic process}, the LLA typically makes almost vanishing {\em all} the terms, appearing in the projection/cumulant expansion, which 
would destroy the FPE.
This means that \textit{often} the LLA FPE would be valid even for large $\epsilon$ 
values for which the cBFPE breaks down 
(but a counterexample is shown if Fig.~\ref{fig:cubica_comparison_cor}).
%

%  Therefore, according to the examples we have illustrated in this paper, although we
%  have obtained the correct BFPE for the SDE of Eq.~(\ref{SDEGen}), given that the
%  LLA FPE is an excellent approximation of the cBFPE
% when the latter is applicable, the simpler analytical expression given by the diffusion coefficient $D(X,t)_{LLA}$ of Eq.~(\ref{DLLAt}) 
% and the typically better performance of the LLA FPE
% compared to the BFPE when $\epsilon$ is increased, we conclude that the LLA FPE is usually preferable.

On the other hand, if the stochastic process $\xi(t)$ is not Gaussian, or it is not 
at all stochastic (for example, it is the degree of freedom 
of a chaotic dynamical system), 
then the Almost Gaussian Assumption or the Fox functional calculus can no longer 
be advocated to give an a priori justification (although weak) to the LLA FPE. 
In these cases, a small $\epsilon$ value and the cBFPE would be the only 
possible approach for a proper FPE treatment, and the LLA FPE could be, at the 
best, an approximation of the cBFPE.
%
%\end{widetext}
%
\section{Conclusions\label{sec:conclusion}}
By definition, the BFPE is the best FPE we can get from a perturbation approach starting from a SDE. In this work we are interested
in the 1-d case with additive noise as in Eq.~(\ref{SDEGen}),
in which  $\epsilon$ is the small parameter. For the 1-d case the BFPE was obtained many years ago by Lopez, West and 
Lindenberg~\cite{lwlPRA37}, but their result reveals unphysical features. In particular, if
 $\tau$ and $\epsilon$ are not fairly small, it may lead to negative values both  of the
diffusion coefficient and of the PDF, in some region of the state space. It is customary to cure this situation by simply
restricting the domain of support of the PDF to exclude these regions. 
It has been argued that this unphysical result of the BFPE might point to problems in the
model used to represent the physical system~\cite{pljrwlPLA136}.
In this work we show, on the contrary, that these problems are due to an incorrect use of the perturbation approach for dissipative systems.
In particular, a proper use of the interaction picture fixes the problem.
The cBFPE gives results that are close to those of numerical simulations of the SDE of Eq.~(\ref{SDEGen}), even
for values of $\epsilon$ and $\tau$ well beyond those allowed by the classical BFPE.  
The stationary PDF is now similar also to that
obtained from the LLA FPE of 
Grigolini~\cite{tgPRA38,ffgmJSP52a} and Fox~\cite{fPRA34}. 

\appendix
\section{The cubic case\label{app:cubic}}

We briefly present the results for the pure cubic case, namely $C(X)= X^3$. This is an extreme non linear case, in fact, also small oscillations are non-linear. 
It is no coincidence that the standard BFPE cannot be used in this case (see below).
 
\begin{figure}[t]
        \centering
        \includegraphics[width=1.0\linewidth]{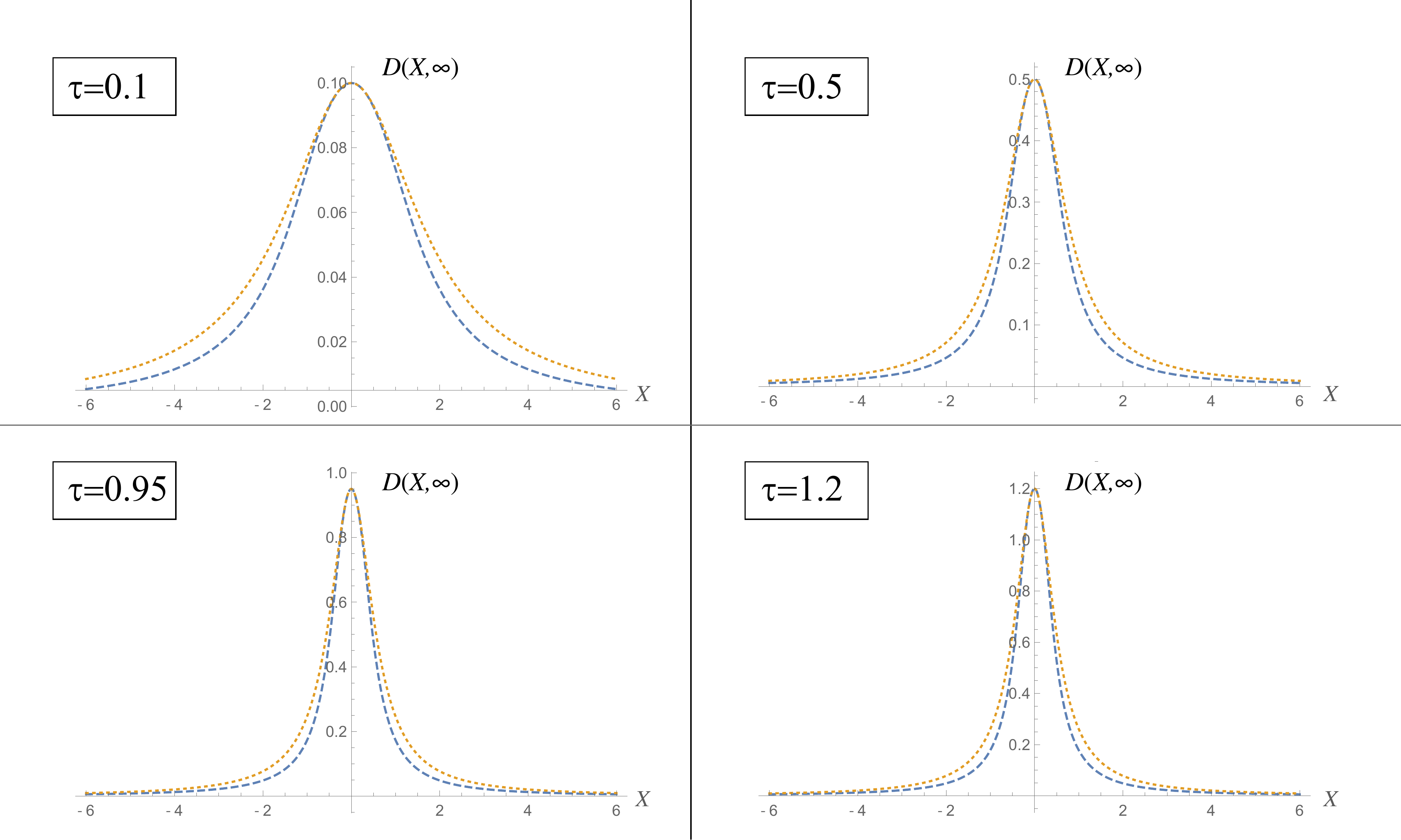}
        \caption{Diffusion coefficients for a pure cubic drift field. The BFPE gives an imaginary result, thus in this case cannot be used. Dashed blue lines: the $D(X,\infty)_{cBFPE}$ of Eq.~(\ref{DcorBFPEcubic}) for different values of $\tau$. 
                Dotted orange line: the $D(X,\infty)_{LLA}$ of Eq.~(\ref{DLLAcubic}) for the same values of $\tau$. }
        \label{fig:Dcubic}
\end{figure}

\begin{figure*}[h]
        \centering4
        \includegraphics[width=1.0\linewidth]{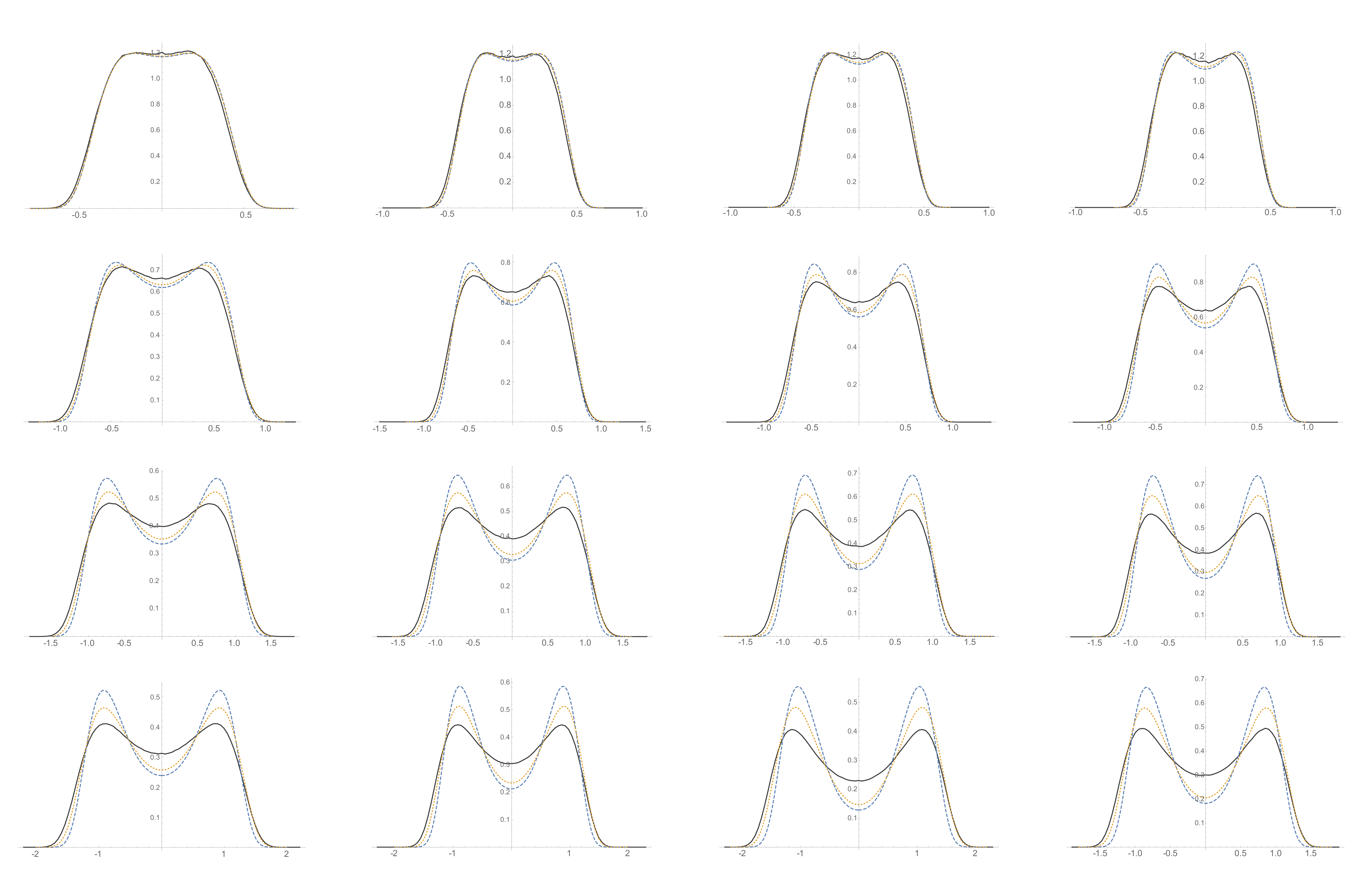}
        \caption{The stationary PDF for the SDE of Eq.~(\ref{SDEGen}) where $C(X)= X^3$      and          $\xi(t)$ is a 
                Gaussian noise with correlation function $\varphi(t)=\exp(-t/\tau)$.      In this case the standard BFPE \textit{cannot}  be used because it leads to an  imaginary diffusion coefficient $D(X,\infty)_{BFPE}$ (see text). The four columns correspond to four 
                different values for $\tau$, while the four rows corresponds to four different values for $\epsilon$. Solid black
                lines: the results of the 
                numerical simulation of the SDE. Dashed blue lines: the corrected BFPE results, namely the PDF of Eq.~(\ref{Ps}) where the diffusion coefficient is given in Eq.~(\ref{DcorBFPEcubic}). 
                Dotted orange lines: the LLA result, namely the PDF of Eq.~(\ref{Ps}) 
where the diffusion coefficient is given in Eq.~(\ref{DLLAcubic}).}
        \label{fig:cubica_comparison_cor}
\end{figure*}

%\begin{widetext}
From Eq.~(\ref{DBFPEt}) we easily obtain,  
\begin{align}
\label{DBFPEcubic}
D(X,t)_{BFPE}=&\epsilon^2\, \frac{1}{2} \,  e^{-\frac{t}{\tau }}
 \bigg[(2 \sqrt{1-2 t X^2} e^{\frac{1}{2 \tau  X^2}} \left(2 t X^2+3 \tau  X^2-1\right)\nonumber \\
 &\left.-3 \sqrt{2 \pi } \tau ^{3/2} X^3 e^{t/\tau }\, \text{erfi}\left(\frac{\sqrt{\frac{1}{2}
                -t X^2}}{\sqrt{\tau } X}\right)\right] e^{-\frac{1}{2 \tau X^2}}\nonumber \\
        &
-\frac{1}{2} \tau  \left[-3 \sqrt{2 \pi } \tau ^{3/2} X^3 e^{-\frac{1}{2 \tau  X^2}}\, \text{erfi}\left(\frac{1}{\sqrt{2} \sqrt{\tau } X}\right)+6 \tau  X^2-2\right]
\end{align}
that, for $t>2 X^2$ \textit{is a complex number}: for large times it is not defined. This means that  for a cubic drift field,  by using the standard BFPE  a stationary PDF cannot be obtained.
%\end{widetext}
The situation is different exploiting our correction to the BFPE. In fact, for large times  ($t\to \infty$), we have (see Eq.~(\ref{Dcor})
\begin{align}
\label{DcorBFPEcubic}
&D(X,\infty)_{cBFPE}=D(X,\bar u(X))_{BFPE}\nonumber \\
&=\epsilon^2 \,
\left[1+ 3 \tau X^2 \left( \sqrt{2}\sqrt{\tau} X F\left(\frac{1}{\sqrt{2} X \sqrt{\tau }}\right)-1\right)\right]
\end{align}
where $F(x):=e^{-x^2} \int_0^x e^{y^2} \, \text{d}y=e^{-x^2} \frac{\sqrt{\pi }}{2}\text{erfi}(x)$
is the Dawson function. The diffusion coefficient of
 Eq.~(\ref{DcorBFPEcubic})
is now positive and well defined for any $X$. 
Concerning the LLA diffusion coefficient, from Eq.~(\ref{DLLA}) we easily get:
\begin{equation}
\label{DLLAcubic}
D(X,\infty)_{LLA}=\frac{\epsilon^2 }{3   \tau  X^2+1}.
\end{equation}
In Fig.~(\ref{fig:Dcubic}) we compare the corrected BFPE and the LLA diffusion coefficients, respectively. Inserting in Eq.~(\ref{Ps}) the expressions in Eqs.~(\ref{DcorBFPEcubic}))-(\ref{DLLAcubic})), we obtain the stationary PDF
shown in Fig.~\ref{fig:cubica_comparison_cor}.
We see that in this extreme non linear case, where the standard BFPE cannot be used, our corrected BFPE gives results that, for small $\epsilon$, are in agreement with
  numerical simulations of the SDE. Notice that, in this case, also the LLA fails for large $\epsilon$ values. 
\bibliography{BiblioCentrale}

\end{document}